\documentclass[%
 aip,
rsi,%
 amsmath,amssymb,
]{revtex4-1}
\usepackage{geometry}
\usepackage{graphicx}
\usepackage{dcolumn}
\usepackage{bm}

\begin{document}


\title{A memory based random walk model to understand diffusion in crowded 
heterogeneous environment}

\author{Sabeeha Hasnain}
\altaffiliation[Present address: ]{Department of Chemistry, University of Texas at Austin, TX 78712, U.S.A.}
\affiliation{School of Computational and Integrative Sciences, Jawaharlal Nehru University, New Delhi 110067, INDIA
}
\author{Upendra Harbola}%
\affiliation{ Department of Inorganic and Physical Chemistry, Indian Institute of Science, Bangalore 560012, INDIA
}%
\author{Pradipta Bandyopadhyay}
\email{praban07@gmail.com}
\affiliation{School of Computational and Integrative Sciences, Jawaharlal Nehru University, New Delhi 110067, INDIA
}


\begin{abstract}
We study memory based random walk models to understand diffusive motion in 
crowded heterogeneous environment. The models considered are non-Markovian as 
the current move of the random walk models is determined by randomly selecting a move from history. At 
each step, particle can take right, left or stay moves which is correlated with 
the randomly selected past step. There is a perfect stay-stay correlation which 
ensures that the particle does not move if the randomly selected past step is a 
stay move. The probability of traversing the same direction as the chosen 
history or reversing it depends on the current time and the time or position of 
the history selected. The time or position dependent biasing in moves implicitly 
corresponds to the heterogeneity of the environment and dictates the long-time 
behavior of the dynamics that can be diffusive, sub or super diffusive. A 
combination of analytical solution and Monte Carlo simulation of different 
random walk models gives rich insight on the effects of correlations on the 
dynamics of a system in heterogeneous environment.
\end{abstract}

\maketitle

%

\section{\label{sec:level1}INTRODUCTION}

Diffusion has always been a subject of interest due to its wide applicability in 
physics, chemistry and biology \cite{frey2005brownian,hasnain2014comparative,george2001transport,kastantin2012identifying,plastino2004effect}. The diffusion of molecules can be 
under the influence of concentration gradient or because of thermal motion of 
the molecules. The diffusive motion of particles can be categorized as normal or 
anomalous depending on the variation of mean square displacement (MSD) with time 
(t). The diffusion is said to be normal when MSD varies linearly with time i.e.  
$MSD \propto t$. However, when
the MSD varies with t as $MSD\propto t^\alpha$($\alpha$ is called the diffusion 
exponent) and $\alpha \neq 1$, diffusion is said to be anomalous. When 
$0<\alpha<1$, the diffusion is said to be subdiffusive and it is superdiffusive 
when $\alpha>1$. 
The subdiffusive dynamics can be due to crowding in a concentrated system which 
can make the system heterogeneous and disordered \cite{ghosh2016anomalous}. The 
crowded environment obstructs the diffusing particle and generally gives rise to 
subdiffusion \cite{vercammen2007measuring,pan2009viscoelasticity,szymanski2009elucidating,ernst2012fractional,jeon2013anomalous,grebenkov2013hydrodynamic,grebenkov2014analytical,ernst2014probing,lee2014characterization,shin2015kinetics}. 
Biological systems are good examples of crowded and 
heterogeneous environments and have been extensively studied 
\cite{hofling2013anomalous,jeon2011vivo,seisenberger2001real,skaug2011correlating,weigel2011ergodic,sahl2010fast,hasnain2014new,mcguffee2010diffusion,nicolau2007sources,janmey1986structure,xie2008single,hammar2012lac}. Experimental studies confirmed the presence of subdiffusion 
while studying the motion of macromolecules inside different biological cells 
\cite{golding2006physical,weiss2004anomalous,banks2005anomalous,kues2001visualization,brown1999measurement,wachsmuth2000anomalous}. However, the 
observed subdiffusion can be a transient one, meaning that the subdiffusion 
$\alpha<1$ becomes normal $\alpha=1$  at long time, or a persistent one, where $\alpha$
always remains less than one.\\
Experimental signatures of both transient and persistent subdiffusion have been 
observed. For instance, in the experimental study by Golding and Cox 
\cite{golding2006physical}, the motion of fluorescently labeled mRNA molecule 
has been tracked inside a live \textit{E. coli} cell and is found to be persistently 
subdiffusive with MSD varying as $MSD \propto t^{0.70}$ . The studies mentioned 
in references \cite{skaug2011correlating,golding2006physical,banks2005anomalous,brown1999measurement,wachsmuth2000anomalous}
 confirm the presence 
of persistent subdiffusion with constant diffusion exponent over all time 
scales. Transient subdiffusion has been observed by Javanainen et al. 
\cite{javanainen2013anomalous} in the study of lateral diffusion of proteins in 
a crowded lipid membrane. Similar results have also been found in references 
\cite{berezhkovskii2014normal,berezhkovskii2014discriminating} and \cite{hasnain2014new}. 
In a recent work \cite{jeon2016protein}, extensive 
molecular dynamics simulations have been performed to determine the effect of 
protein crowding on membrane dynamics. The simulation study of lipids in the 
presence of protein or cholesterol as crowding particles shows persistent 
anomalous subdiffusion dynamics for both lipids and membrane-embedded proteins, 
which is governed by a non-Gaussian distribution \cite{metzler2016non}.\\
Theoretical models like Fractional Brownian motion 
\cite{mandelbrot1968fractional}, Continuous time random walk, 
\cite{metzler2000random} and Obstructed diffusion \cite{havlin1987diffusion} 
have been utilized by previous studies to understand subdiffusion in crowded 
environment. Mandelbrot and Van Ness \cite{mandelbrot1968fractional} showed that 
when the direction of motion of a particle is determined from the history in a 
power law fashion, which can be either correlated or anti-correlated, diffusion 
is found to be anomalous and is termed as Fractional Brownian Motion. The origin 
of anomaly in this case is long-range temporal rather than spatial correlation. 
Power laws occur frequently in the diffusion in heterogeneous environments with 
multi-scale features but differing in their origin 
\cite{fish2010multiscale,coppens2006effects}. Previously Hasnain et al. 
\cite{hasnain2014new} found transient subdiffusion for protein diffusion in a 
cytoplasm. The random walk model described in reference \cite{hasnain2015analytical} 
is appropriate for transient sub-diffusion in 
crowded environment but not for describing persistent sub-diffusion.\\
In the current work, we study microscopic random walk models to describe 
persistent subdiffusion in heterogeneous environment. The main motivation behind 
our study is to incorporate effects of dynamic heterogeneity to the existing 
model by introducing dynamic correlations between the current step and the 
history. Our starting point is the model developed by Kumar et al. 
\cite{kumar2010memory}(henceforth this model will be referred as Kumar's model). In that model, the authors developed a memory based 
random walk model in which the current step depends on the randomly selected 
past step. At each step, the particle can take one of the three steps; left, 
right and stay (i.e. does not move). In the model, the stay moves are perfectly 
correlated which implies that if the past step selected is a stay move, then the 
particle will stay at its position with probability one. However, if the past 
step selected is right (left), the particle has the probability to take right 
(left) move with probability `p' or chooses to reverse its direction with 
probability `q'. It can also stay at its position with probability `r'. The 
parameters `p' and `q' are taken as constants and are independent of the current 
step and the past step selected. The model can describe all types of diffusion, 
namely superdiffusion, normal and subdiffusion. In a similar work by Harbola et 
al. \cite{harbola2014memory}, the authors proposed a minimal- option model for 
the walker. A walker can take either forward or stay move with perfect 
correlation in the stay moves. The model also shows all types of diffusion such 
as subdiffusive, superdiffusive and normal. However, the random walk models 
discussed above \cite{kumar2010memory,harbola2014memory} do not account for the 
heterogeneity of the environment and its effect on the dynamics of particle. In 
the present work, we show that the heterogeneity of the environment can give 
rise to qualitative changes in dynamics which has not been discussed in previous 
literature.\\
In the current work, we have implicitly included the effects of heterogeneity of 
the system and crowding on the dynamics of diffusing particle both in an average 
manner and as local crowding. First, the average crowding in the environment has 
been included by allowing the particle to stay at the current position with 
probability (r) which is the probability of the occupancy of the neighboring 
lattice sites. Secondly, the probabilities (p and q) to choose the direction of 
the next step are considered as functions of the current and the past times and positions (i.e. there are two different models for temporal and spatial 
dependence). This tries to take care of local (dynamic) crowding since the 
presence of local heterogeneity in the system may lead to spatial and temporal 
correlation between the past and present moves. The randomly chosen steps from 
immediate past are sometimes followed with lower probability than those chosen 
from the distant past and vice versa. The current model differs from Kumar's 
model in the sense that, in the present model, the environment heterogeneity 
dynamically influences the efficiency with which the particle follows a past 
step. As we discuss below, this heterogeneity effect leads to qualitative 
changes in the dynamics predicted by the Kumar's model.\\
One of the models proposed here give all three types of diffusion but the other 
two models give only subdiffusion. Hence, the dynamical behaviors depend on the 
type of correlation induced by heterogeneity.
The paper is organized in the following manner. Methodology section describes 
Kumar's model and our extension of it. In the method section, we have also given 
analytical formulation and Monte Carlo simulation schemes performed. Result 
section gives the features of the models using MSD, diffusion exponent ( 
$\alpha$) and probability distribution function (PDF). A summary of the three 
models and their connection to the heterogeneity of the environment has also 
been discussed. The paper ends with a conclusion and possible future work.\\
\section{\label{sec:level1}METHODOLOGY}
Several random walk models have been proposed to understand the mechanism of 
subdiffusion in crowded environment 
\cite{metzler2000random,bouchaud1990anomalous,ben2000diffusion,metzler2014anomalous,meroz2015toolbox,saxton1997single,vilaseca2011new,isvoran2008computational}. 
A simple random walk consists of a series of right 
and left moves along a one-dimensional lattice 
\cite{chandrasekhar1943stochastic} with equal probability which is independent 
of the previous steps taken. This type of motion with independent steps gives 
rise to normal diffusion at long times where MSD varies linearly with time. 
However, when the walk is biased in a direction, leading to drift in that 
direction, it is said to be a biased random walk \cite{berg1993random}. If the 
steps are correlated it is called a correlated random walk 
\cite{montroll1965random,konno2009limit} which may give rise to
anomalous diffusion, i.e., subdiffusion and superdiffusion. 
Several theoretical and computational 
models have been developed in the past which can produce transient subdiffusion 
\cite{hasnain2014new,berezhkovskii2014normal,berezhkovskii2014discriminating,ando2010crowding,ridgway2008coarse} . 
However, only few microscopic models are 
known \cite{kumar2010memory,harbola2014memory} to explain normal diffusion, 
persistent subdiffusion and superdiffusion within the same scheme.

\subsection{\label{sec:level2}Kumar's Model}
This model consists of a random walker moving on a one-dimensional infinite 
lattice where the lattice points are unit distance apart. The starting step 
($\sigma_1$) is selected in the right or left direction with probability s or 
(1-s), respectively where $s > 0$ . The subsequent steps can be right, left or stay 
which is decided as the following.  At each step, a past step is selected 
uniformly from the history which decides the current move of the particle. If 
the past step selected is a stay move, then the particle remains at the present 
position with probability 1. However, if the past step selected is right or 
left, then the particle has the tendency to move in the same or reverse 
direction with probability p and q, respectively. It can also stay in the same 
point with probability r. In this model p is said to be the probability of going 
in the same direction and q is the probability of reversing the direction and 
these values are taken as constants, independent of the current and past steps. 
At each step, the sum of p, q and r should be equal to 1. The model gives 
subdiffusion, superdiffusion and normal diffusion depending on the asymmetry 
parameter $\gamma$ where $\gamma=p-q$ .
The position $x_{n+1}$ of the particle at step n+1 is given as ( $x_n$ is the 
position after step n)
\begin{eqnarray}
\label{1inold}
x_{n+1}=x_n+\sigma_{n+1}
\end{eqnarray}
 where $\sigma_{n+1}=\pm 1,0$ is the current move at ${n+1}^{th}$ step which is 
decided from a randomly selected past step from the history $\left\{\sigma_1, 
\sigma_2, \sigma_3,....,\sigma_n\right\}$ with uniform probability 1/n.
For the first step,
$\sigma_1=\pm 1$\\
If $\sigma_k$ is the randomly selected past step, then\\
$\sigma_{n+1}=\sigma_k$ with probability p\\
$\sigma_{n+1}=-\sigma_k$ with probability q\\
$\sigma_{n+1}=0$    with probability r\\
It is crucial to have perfect correlation between the stay moves, otherwise only 
normal and transient subdiffusive or superdiffusive dynamics is predicted by 
this model \cite{harbola2014memory}.\\
The random walk with the given probabilities can be described as the following. 
For the first step at time t=1, the probability that $\sigma_1=\sigma$ is given 
by\\
\begin{eqnarray}
\label{P1sigma}
P[\sigma_1=\sigma]=\frac{1}{2}[1+(2s-1)\sigma] \,,  \mbox{where}\,  \sigma=\pm 1
\end{eqnarray}
For time t+1 ($t \geq 1$), the conditional probability to make a move $\sigma(=1,-1,0)$ 
is given as
\begin{eqnarray}
\label{Psigma}
P[\sigma_{t+1}=\sigma|{\sigma_t}]=1-\sigma^2 +\frac{1}{2t} \sum_{k=1}^t 
\sigma_k^2(3\sigma^2-2)(1-r)+\sigma \sigma_k \gamma
\end{eqnarray}
Here $\gamma=p-q$ . Using Eq. \ref{Psigma}, the first two moments of the 
displacement after time t can be obtained as shown in previous works 
\cite{kumar2010memory,harbola2014memory}.
\subsection{\label{sec:level2}Extension of Kumar's model to incorporate environmental 
heterogeneity}
In the current work, we are proposing a model to understand anomalous diffusion 
in complex heterogeneous environment. In our model, we associate the stay 
probability (r) to the occupancy of the lattice sites i.e. the fractional volume 
occupancy of a crowded system. This implicitly includes the effect of crowding 
in an average manner. In Kumar's model p and q are constants which do not 
describe the heterogeneity of the environment of the diffusing particle. To 
account for the heterogeneity of the environment, we consider p  and q as 
functions of current time t, and the history selected. Note that the time t is 
analogous to the step number. The time dependence of p and q accounts for the 
local heterogeneity of the system. We have kept r fixed for a particular study 
so only one independent parameter p (or q) is required to specify the model. For 
the current study we have taken three different cases for the selection of 
probability p which are given below
\subsubsection{Model 1:}
\begin{eqnarray}
\label{EQmod1}
p(t,k)={(\frac{t-k}{t})}^\beta (1-r) \, \mbox{,where} \, \beta>0
\end{eqnarray}
where $t$ is the current time, $k$ is the time of randomly selected past step, r is 
the stay probability and $\beta$ is a parameter determining the heterogeneity of 
the environment. From the above expression, it can be said that $p(t,k)$ will have 
higher value when the selected step k is far from the present time compared to 
the case where k is close to the present time. The parameter $\beta$ determines 
how efficiently particle follows the past i.e. in this case the far history is 
followed with more probability than the close one. 
\subsubsection{Model 2:}
\begin{eqnarray}
\label{EQmod2}
p(t,k)={\exp{(-(t-k)^2)}}(1-r)
\end{eqnarray}
This represents a Gaussian time-correlation between present and past selected 
moves. From the above expression, it can be said that $p(t,k)$ with a randomly 
selected step closer to the present time has higher value than the other case.
\subsubsection{Model 3:}
\begin{eqnarray}
\label{EQmod3}
p(t,k)=(1+(t-k)^2)^{-\epsilon}(1-r)
\end{eqnarray}
where $\epsilon$ is a parameter having value greater than zero. This model also 
shows that $p(t,k)$ will have larger value for randomly selected step close to the 
current state.\\
For the above three models, $p(t,k)$ is a function of time only and hence we call it 
as temporal dependence henceforth. However, we shall also analyze the cases when 
probabilities of taking moves are functions of positions at times t and k . We 
shall call this as spatial dependence. For brevity, henceforth we shall write 
$p(t)$ (or $p(x)$, when it is a function of position) as p only.\\
Let $p(t,k) [q(k,t)]$ be the probability to follow [reverse] a randomly chosen 
$k^{th}$ step
at time $t$, if $\sigma_k=\pm 1$. Then the conditional probability, 
$P[\sigma_{t+1}=+1|\{\sigma_t\}]$, 
to have step $\sigma_{t+1}=+1$ for a given history $\{\sigma_{t}\}$ can be 
written as,
\begin{eqnarray}
 \label{eq-1}
 P[\sigma_{t+1}=+1|\{\sigma_t\}] = 
\frac{1}{t}\sum_{k=1}^t\left[\delta_{\sigma_k,+1}p(t+1,k)
 +\delta_{\sigma_k,-1}q(t+1,k)\right]
\end{eqnarray}
where $\delta_{\sigma_k,\pm 1}$ is the kronecker
 delta function between $\sigma_k$ 
and $\pm 1$. 
Since $\sigma_k=\pm 1$ or $0$, we can re-express the above equation in terms of 
$\sigma_k$ as,
\begin{eqnarray}
 \label{eq-2}
 P[\sigma_{t+1}=+1|\{\sigma_t\}] = 
\frac{1}{2t}\sum_{k=1}^t\sigma_k \left[(1+\sigma_k)p(t+1,k)
 -\sigma_k(1-\sigma_k)q(t+1,k)\right]
\end{eqnarray}
This can be rearranged to
\begin{eqnarray}
 \label{eq-3}
 P[\sigma_{t+1}=+1|\{\sigma_t\}] = \frac{1}{2t}\sum_{k=1}^t\left(\frac{}{}\sigma_k^2[p(t+1,k)+q(t+1,k)]
 +\sigma_k[p(t+1,k)- q(t+1,k)]\frac{}{}\right)
\end{eqnarray}
Similarly, for $P[\sigma_{t+1}=-1|\{\sigma_t\}]$, we obtain,
\begin{eqnarray}
 \label{eq-4}
 P[\sigma_{t+1}=-1|\{\sigma_t\}] = 
\frac{1}{2t}\sum_{k=1}^t\left(\frac{}{}\sigma_k^2[p(t+1,k)+q(t+1,k)]
 -\sigma_k[p(t+1,k)- q(t+1,k)]\frac{}{}\right)
\end{eqnarray}
Since $P[\sigma_{t+1}=0|\{\sigma_t\}]=1- 
P[\sigma_{t+1}=+1|\{\sigma_t\}]-P[\sigma_{t+1}=-1|\{\sigma_t\}]$,
one obtains,
\begin{eqnarray}
 \label{eq-5}
 P[\sigma_{t+1}=0|\{\sigma_t\}] = 1-\frac{1-r}{t}\sum_{k=1}^t\sigma_k^2.
\end{eqnarray}
We can combine Eqs. (\ref{eq-3})-(\ref{eq-5}) in to a single equation as,
\begin{eqnarray}
 \label{eq-6}
 P[\sigma_{t+1}=\sigma|\{\sigma_t\}] = 1-\sigma^2+\frac{1-r}{2t}\sum_{k=1}^t 
 \left[\frac{}{} (3\sigma^2-2)\sigma_k^2  +\frac{\sigma \sigma_k}{1-r} 
(2p(t+1,k)- 1+r)\frac{}{}\right]
\end{eqnarray}
where $\sigma=0,\pm 1$.
Several things can be derived starting from Eq. (\ref{eq-6}). Let $p_{\pm}(t)$ 
be the probability
of the $t^{th}$ step to be $\pm 1$, and similarly $p_0(t)$ for the $t^{th}$ step to be 
zero. 
These probabilities are then
obtained by averaging Eq. (\ref{eq-6}) over all histories. For example,
\begin{eqnarray}
 \label{eq-7}
 p_{\pm 1}(t+1) = \frac{1-r}{2t}\sum_{k=1}^t 
 \left[\frac{}{}\langle \sigma_k^2\rangle  
 +\frac{\sigma \langle \sigma_k\rangle }{1-r} (2p(t+1,k)- 1+r)\frac{}{}\right]
\end{eqnarray}
Note that averages, $\langle \sigma_k\rangle$ and $\langle \sigma_k^2\rangle$ 
can be expressed
in terms of $p_{\pm}(k)$ as $\langle\sigma_k\rangle=p_{+}(k)-p_{-}(k)$ and
$\langle\sigma_k^2\rangle=p_{+}(k)+p_{-}(k)$. Using this in Eq. (\ref{eq-7}) and 
the fact
that $p_0(t)=1-p_+(t)-p_-(t)$, we obtain a recursive relation for $p_0(t)$,
\begin{eqnarray}
 \label{eq-8}
 p_0(t) = \frac{r}{t-1} +\frac{t-r-1}{t-1}p_0(t-1).
\end{eqnarray}
It can be solved to obtain,
\begin{eqnarray}
 \label{eq-9}
p_0(t)=1-\frac{\Gamma(t-r)}{\Gamma(t)\Gamma(1-r)}
\end{eqnarray}
where $\Gamma$ refers to the gamma function and $t>1$. This immediately gives,
\begin{eqnarray}
 \label{eq-10}
p_+(t)+p_-(t)=\frac{\Gamma(t-r)}{\Gamma(t)\Gamma(1-r)}
\end{eqnarray}
when $t>1$. 
We next start from Eq. (\ref{eq-7}) to calculate $\Delta 
p(t)=p_+(t)-p_-(t)$. We get,
\begin{eqnarray}
 \label{eq-11}
 \Delta p(t+1) = \frac{1}{t}\left[2p(t+1,t)+r+t-2\right]\Delta p(t) 
 +\frac{2}{t} \sum_{k=1}^{t-1} \Delta p(k)[p(t+1,k)-p(t,k)].
\end{eqnarray}
Since $\Delta p(1)=2s-1$, from Eq. (\ref{eq-11}), all $\Delta p(k)$ are 
proportional
to $2s-1$. Thus for $s=1/2$, $\Delta p(k)=0 ~ \forall k>0$. Thus for $s=1/2$,
\begin{eqnarray}
 \label{eq-12}
p_+(t)=p_-(t) = \frac{\Gamma(t-r)}{2\Gamma(t)\Gamma(1-r)}.
\end{eqnarray}
Indeed this immediately leads to 
\begin{eqnarray}
 \label{eq-13}
 \langle \sigma_k\rangle &=& 0\\
 \langle \sigma_k^2\rangle &=& \frac{\Gamma(k-r)}{\Gamma(k)\Gamma(1-r)} 
\end{eqnarray}
and therefore $\langle x_t\rangle =0$ $\forall$ t.
Because of the complexity of the expressions, it has not been possible to derive the expression for the second moment of displacement. The second moment of displacement has been calculated numerically using Monte Carlo simulation scheme.
\subsection{\label{sec:level2}Monte Carlo Simulations}
Because of the complexity of the models in the current work, we have used Monte Carlo (MC) simulations to the dynamic behavior for different models as given in Eq. 
\ref{EQmod1}-\ref{EQmod3}. For each model, corresponding to each walk length, we 
have run 9 million MC simulations. The MSD and diffusion exponent ($\alpha$) 
have been calculated for each model. From each case, our focus is on 
understanding the diffusive behavior at different values of volume occupancy 
(given by r) and the environmental heterogeneity (given by the time or spatial 
dependence of p and q). For comparison, we have run MC simulations for Kumar's 
model with the appropriate parameters. For $p-q<0$, Kumar's model 
\cite{kumar2010memory} always gives rise to subdiffusion with $\alpha=1-r$. For 
comparison, we took p=0.2 and q=0.6 and the stay probability, r=0.2 for Kumar's 
model to compare the models developed in our work (where stay probability r is 
taken as 0.2). Our models’ p and q are determined from Eqs. 
\ref{EQmod1}-\ref{EQmod3} corresponding to model 1, 2 and 3 respectively.\\
In the next section, we discuss simulation results for each 
model and make comparison with Kumar's model wherever possible.

\section{\label{sec:level1}RESULTS}
\subsection{\label{sec:level2}Model 1}
\begin{figure}[htbp!]
\includegraphics[scale=0.7]{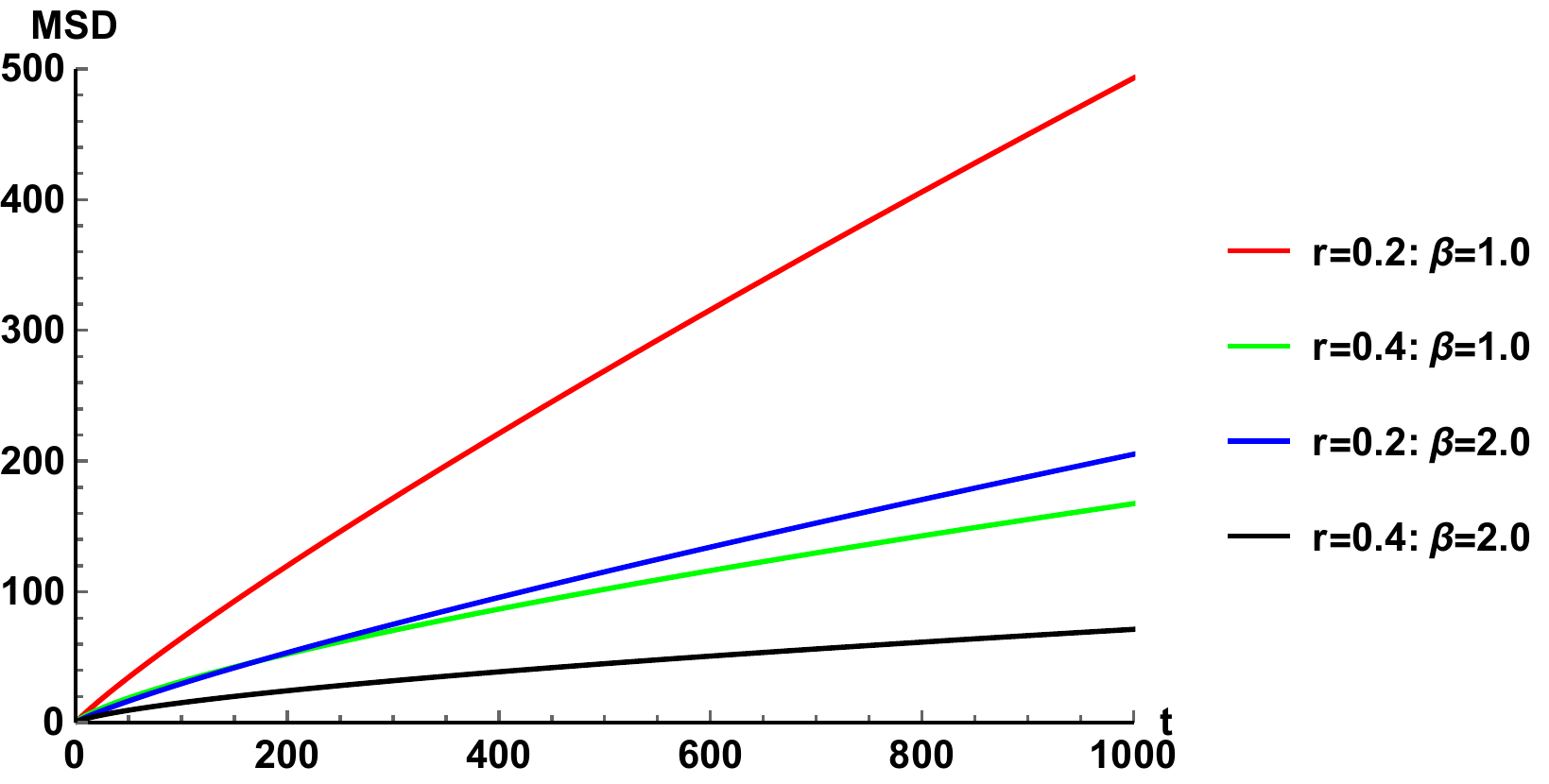}
 \caption{Figure shows MSD, obtained from model 1, plotted against time (t) at stay probability (r=0.2, 0.4) and heterogeneity parameter 
($\beta=1.0, 2.0$).}
  \label{msdfig1}
\end{figure}

For model 1, we have calculated the second moment of displacement for different values of stay probability `r' and heterogeneity parameter `$\beta$'.  Figure \ref{msdfig1} shows mean square displacement (MSD) plotted againt N for stay probability r=0.2, 0.4 and heterogeneity parameter $\beta=1.0.,  2.0$. Figure shows decrease in the value of MSD with increase in heterogeneity parameter and stay probability. With increase in the value of $\beta$, the probability of reversing the history increases with leads to the decrease in the value of MSD. 

\begin{figure}[htbp!]
\includegraphics[scale=1.0]{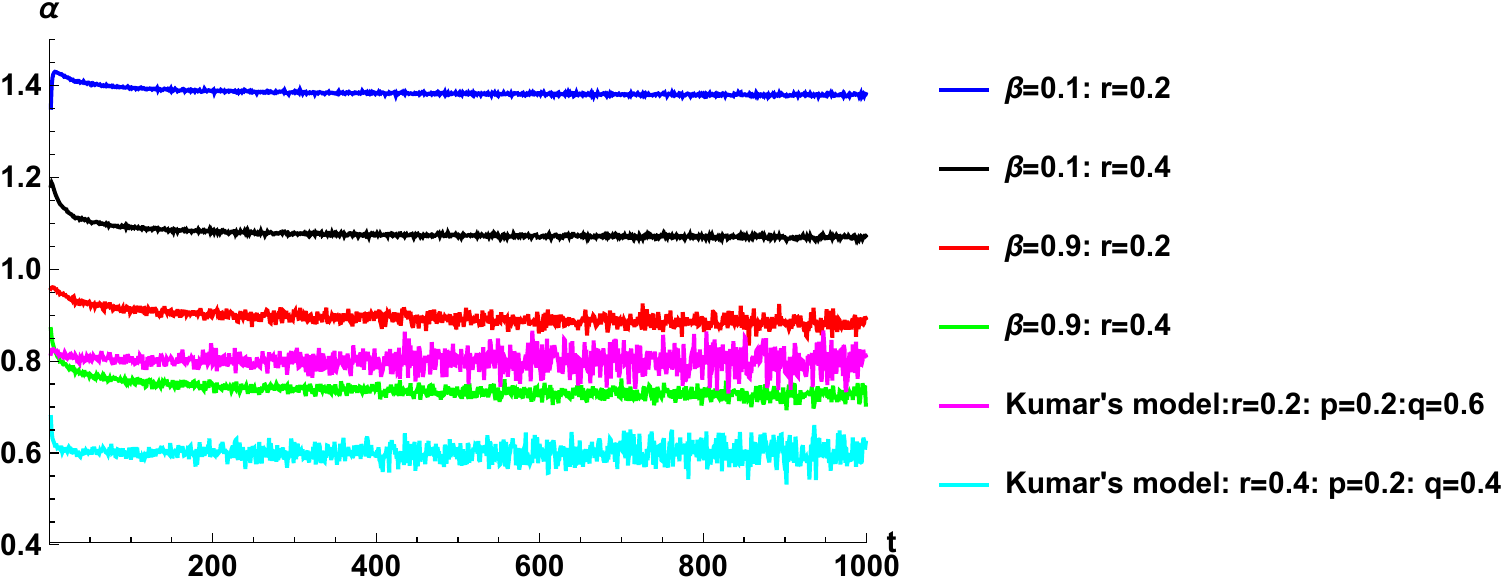}
 \caption{Figure shows diffusion exponent ($\alpha$), for model 1, plotted against time (t) 
for different values of stay probability (r) and heterogeneity parameter 
($\beta$).}
  \label{fig3}
\end{figure}

In Fig. \ref{fig3}, we show the diffusion exponent ($\alpha$) plotted against 
time (t) for r=0.2, 0.4 and $\beta=0.1, 0.9$. We observe that, initially, the 
diffusion exponent decreases with increase in t (except for the blue curve, 
which increases at short time), until it converges to some constant value. For 
r=0.2,
superdiffusive motion (with $\alpha=1.38$) is observed at $\beta=0.1$ and is 
subdiffusive (with $\alpha=0.88$) at $\beta=0.9$. Similar qualitative behavior 
is observed for r=0.4, that is superdiffusive (with $\alpha=1.08$) at 
$\beta=0.1$ and is subdiffusive (with $\alpha=0.74$) at $\beta=0.9$. \\
For a given value of r the dynamics changes significantly with change in value 
of $\beta$ . The change in $\beta$ , implicitly representing the heterogeneity 
of the environment, is leading to qualitative changes in dynamics. Figure 
\ref{fig3} also shows comparison of the current model with Kumar's model for two 
values of stay probability. From Kumar's model, for r=0.2, persistent 
subdiffusion is observed with exponent $\alpha=1-r=0.8$ when $p-q<0$. From the 
current model, at r=0.2 we get superdiffusion, normal (not shown in the figure) 
or subdiffusion with exponent depending on value of $\beta$ which incorporates 
effect of environment. Similarly, for r=0.4, we see a qualitative effect of 
heterogeneity which changes the long-time dynamics from superdiffusion, at 
$\beta = 0.1$, to subdiffusion, at $\beta=0.9$, and differentiates the dynamics 
from Kumar's model 
\cite{kumar2010memory}.\\
\begin{figure}[htbp!]
\includegraphics[scale=0.7]{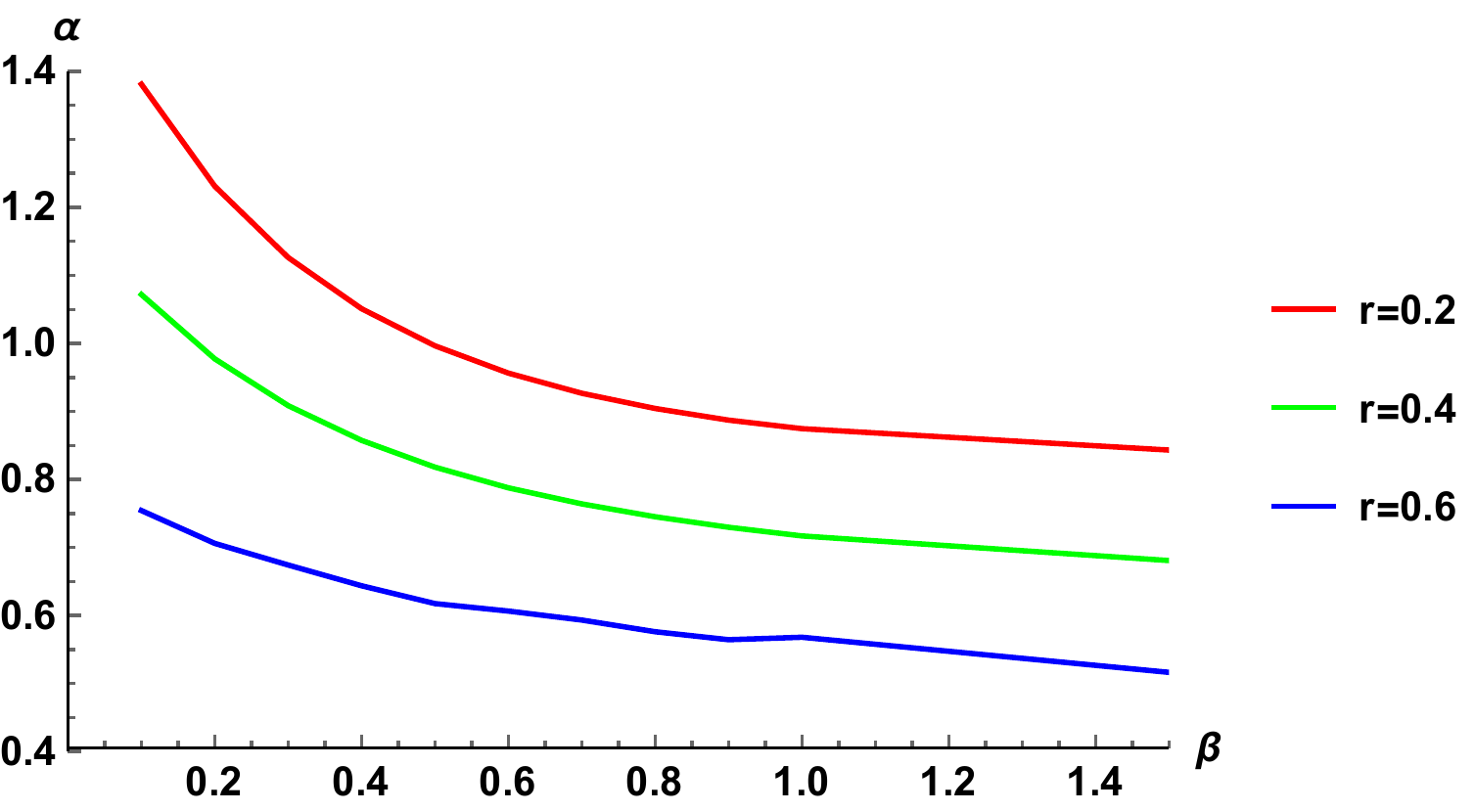}
 \caption{Figure shows variation of diffusion exponent ($\alpha$) for model 1 plotted against heterogeneity parameter ($\beta$) for different values of stay probability (r).
}
  \label{fig4}
\end{figure}
\begin{figure}[htbp!]
\includegraphics[scale=0.9]{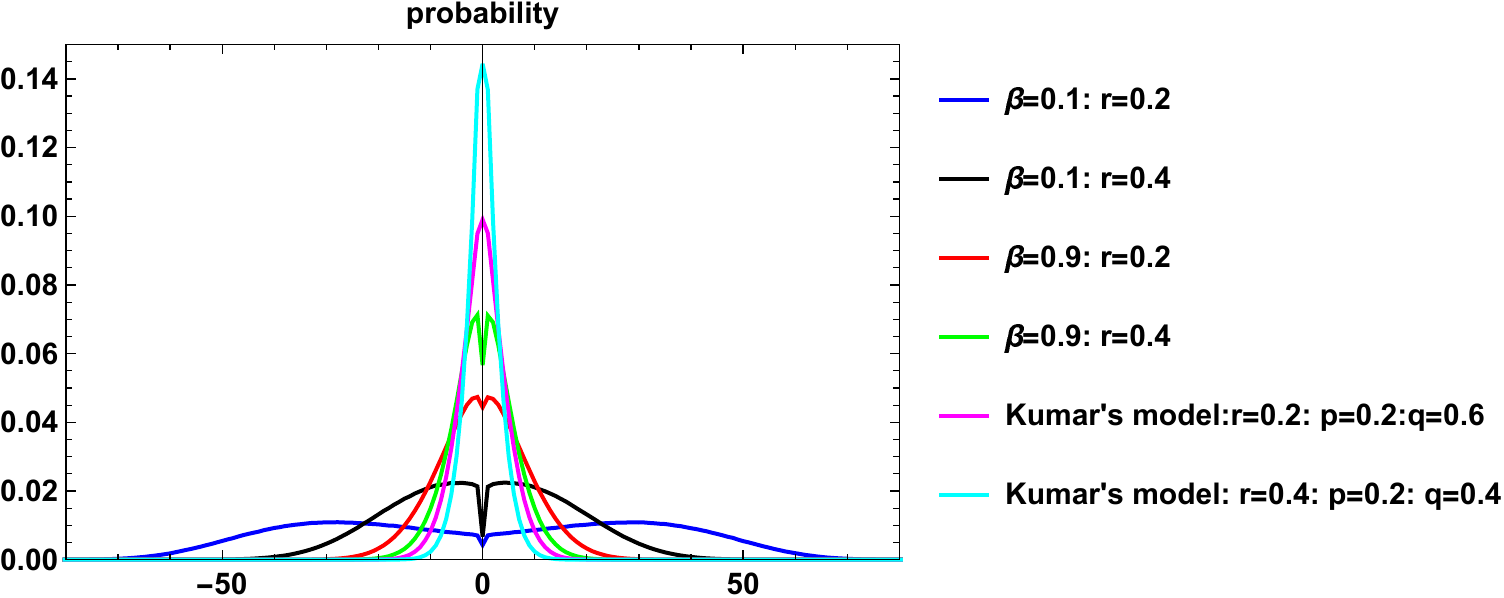}
 \caption{Figure shows probability distribution function for walk of length t=100 steps at different values of 
stay probability (r) and heterogeneity parameter ($\beta$) obtained under model 1.}
  \label{fig5}
\end{figure}
In Fig. \ref{fig4}, we show simulation results for diffusion exponent ($\alpha$) 
plotted against $\beta$ for different values of r. The exponent $\alpha$ 
decreases with increase in $\beta$ . For small values of $\beta$ , $\alpha$ 
decreases sharply and then gradually settles to a constant value. For r=0.2 and 
r=0.4, we see qualitative change in the dynamics with increase in the value of 
beta. For smaller values of $\beta$ motion is superdiffusive and it goes to 
subdiffusive as $\beta$ value increases. For any given value of volume occupancy 
(r), increase in the parameter $\beta$ leads to a decrease in `p' and 
consequently increase in the value of `q'. The increase in the value of `q' 
allows the particle to reverse its direction more for any chosen history, which 
can shift the qualitative behavior of diffusion from superdiffusive to 
subdiffusive as shown in figure \ref{fig4}. However, for large volume occupancy 
r=0.6, we observe subdiffusive motion for all values of $\beta$.
We next look at the heterogeneity effects on the full probability distribution 
of position of the walker. In Fig. \ref{fig5}, we show PDF for 
different values of stay probability (r) and heterogeneity parameter ($\beta$) 
for walks of length up to 100 steps obtained from MC simulations. The 
distribution is symmetric around the origin with two peaks on each side of the 
origin. The symmetry is due to the choice $s=1/2$ which implies that the probabiility of the first step is taken as $1/2$ in both right and left direction. With the increase in the stay 
probability r, the distribution becomes more and more peaked around the origin 
with two peaks getting closer to each other, and the dip at the origin becomes 
deeper. An increase in $\beta$ also makes the distribution more confined around 
the origin. This is understandable as diffusion becomes slower ($\alpha$ 
decreases) with increase in r and $\beta$. To understand the dip at x=0 we 
consider the extreme case when $\beta=0$ which gives $p(t)=(1-r)$ i.e. 
$q(t)=0.0$ (see Eq \ref{EQmod1}). This makes the particle to move in the 
direction of the first step which is always away from the origin ($\sigma_1=\pm 
1$), giving zero probability for particle to be at x=0. With increase in the 
value of $\beta$, the probability of reversing direction increases 
in time which increases the probability at position x=0 but is always less than its 
neighboring positions which are more probable. The figure also shows PDF obtained 
from Kumar's model at r=0.2 and 0.4 at p=0.2. From the figure, we see that for a 
given value of `r', the distributions obtained from Kumar's model are more 
peaked than the one obtained from our model at different values of parameter 
$\beta$. The difference in PDF is due to the change in the value of `p' due to 
the heterogeneity parameter in our model, unlike Kumar's model which has a constant 
value of `p'.
\subsection{\label{sec:level2}Model 2}
\begin{figure}[htbp!]
\includegraphics[scale=1.0]{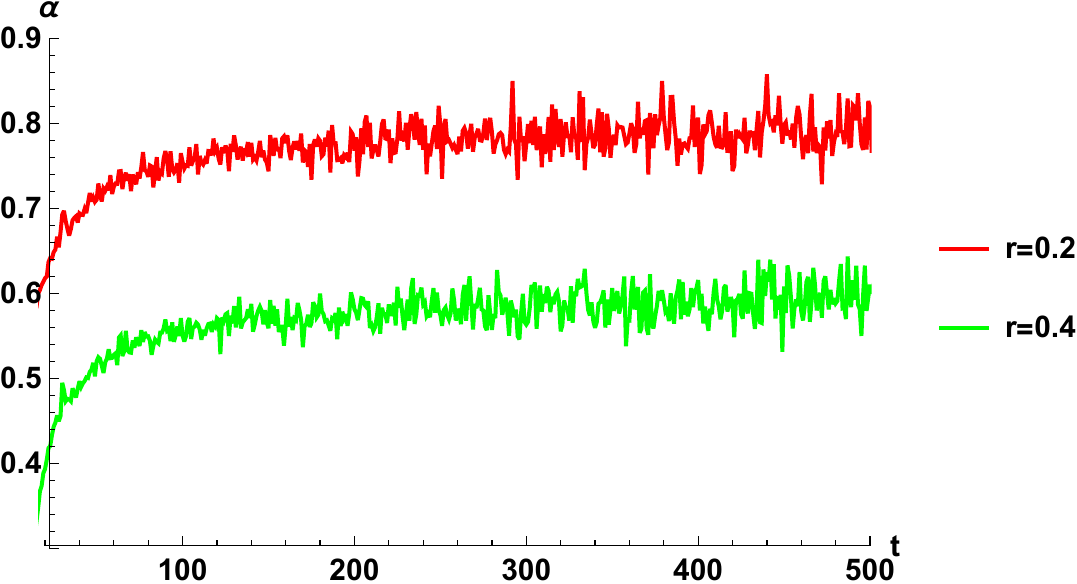}
 \caption{Figure shows variation of diffusion exponent ($\alpha$) for model 2 plotted 
against time (t) for different values of stay probability (r).}
  \label{fig6}
\end{figure}

The MC simulations have been performed using Eq.\ref{EQmod2} as the probability 
of following a randomly selected past step. The probability of following or 
reversing the selected history depends on the current time (t) and the history 
selected ( k ). Figure \ref{fig6} shows diffusion exponent ($\alpha$) , obtained 
from MC simulations, plotted against time (t) at r=0.2 and r=0.4 . From the 
figure, we see that the diffusion exponent ($\alpha$) in each cases converges to 
1-r with increase in time `t'. The dynamics for this model is similar to Kumar's 
model with $p-q <0.5$ and $r<1-2(p-q)$ where the motion is subdiffusive with 
exponent $\alpha=1-r$. With increase in time, the difference $t-k$ increases, 
which causes decrease in value of `p'. The value of `p' at somes point becomes 
negligible in comparison to q (=1-p-r) which corresponds to subdiffusive ($\propto 
t^{(1-r)}$) kind of dynamics mentioned in reference \cite{kumar2010memory}.
Using MC simulations, we also study the case when the probability of following a 
selected history, p, is a function of the current position and the position at 
the randomly selected history. The walk with space dependent probability, p(x), 
is termed here as spatially correlated walk. The spatial correlation was not 
considered for model 1 because the function p(x) is not defined if the particle 
is at x=0 at any time.
For spatially correlated walk, the probability of following the selected past 
step is defined as \\
\begin{eqnarray}
\label{old33}
p=\exp(-(x(t-1)-x(k))^2)(1-r)
\end{eqnarray}
where $x(t)$ is the position at time t. Note that with this form, p becomes more 
fluctuating quantity than its
temporal counterpart.
\begin{figure}[htbp!]
\includegraphics[scale=1.0]{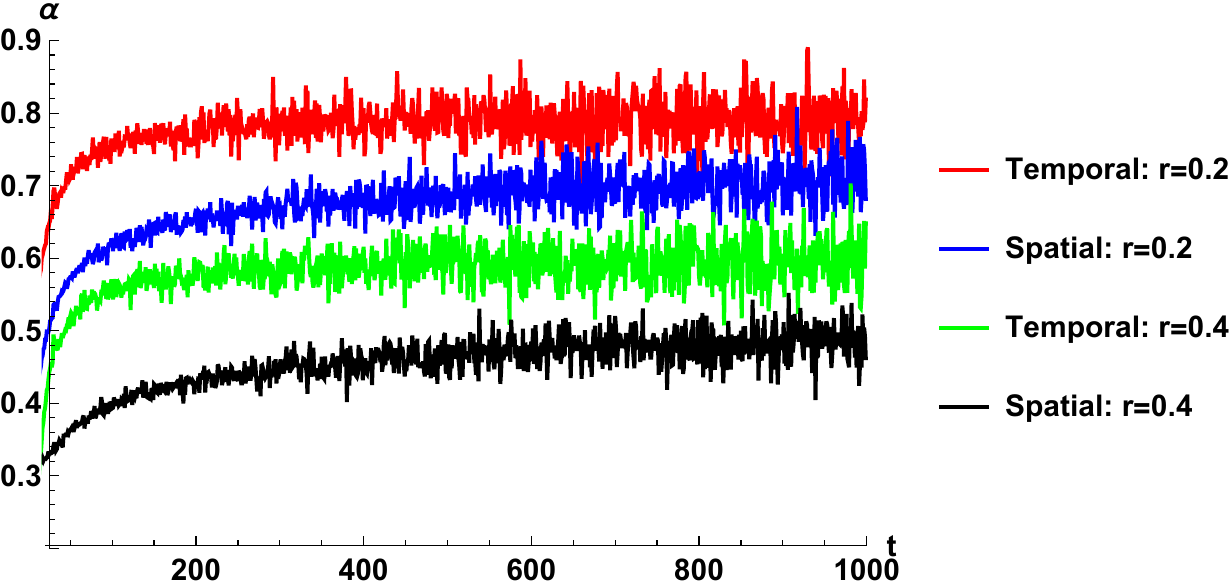}
 \caption{Figure shows variation of diffusion exponent ($\alpha$) for spatial 
and temporal correlated walks under model 2 plotted against time (t) for different values of 
stay probability (r).}
  \label{fig7}
\end{figure}
Figure \ref{fig7} shows time dependence of the diffusion exponent for the 
spatially and the temporally correlated dynamics at r=0.2 and r=0.4. For both 
spatial and temporal correlations, the increase in volume occupancy `r' gives 
rise to more subdiffusive behavior with smaller value of diffusion exponent 
$\alpha$ as shown in the figure. For spatial dependence, diffusion exponent is 
found to be lower than temporal dependence. At any time, t, x(t)-x(k) is less 
than or equal to t-k which makes the probability `p' for spatially correlated 
walk to be larger than temporally correlated one, $p(x(t)) > p(t)$, the effect 
of which is observed in simulations. For temporal correlated walk, diffusion 
exponent $\alpha$ converges to $1-r$ at long time unlike the spatial correlated 
walk which converges to lower value. The exponent in case of spatial correlated 
walk is not just dependent on r but also the values of p and q.
Figure \ref{fig8} compares probability distribution function for the spatially 
and the temporally correlated walks. For points farther from the origin, 
probability for spatial dependent walk is more than that of temporal dependent 
walk. Since $p(x(t))>p(t)$, this may account for the higher probability for 
points far from the origin. On the other hand, probability conservation requires 
that the points close to the origin have comparatively less probability, as seen 
in the figure. The probability of finding the walker at any position x(t) at 
time t depends both on number of paths leading to that position and the 
probability of each path. For position x=0, the number of paths are always 
larger than its neighboring position but the sum of probabilities of paths is 
less than those of the neighboring positions, which leads to a dip at x=0. 
Figure also shows PDF obtained from Kumar's model at r=0.2 and r=.4 and p=0.2. 
For temporal correlated walk (in each case r=0.2 and r=0.4), the PDF is closer 
to the one obtained from Kumar's model but have higher peak at the mean position 
and less displacement around the mean position. The difference in the 
distribution is due to the difference in `p' values in the current model and in 
Kumar's model. In the current model, unlike Kumar's model `p' changes at each 
step and is a function of current step and the history selected which in general 
decreases with time leading to more direction reversal and more confined motion 
around the mean position. 
\begin{figure}[htbp!]
\includegraphics[scale=0.7]{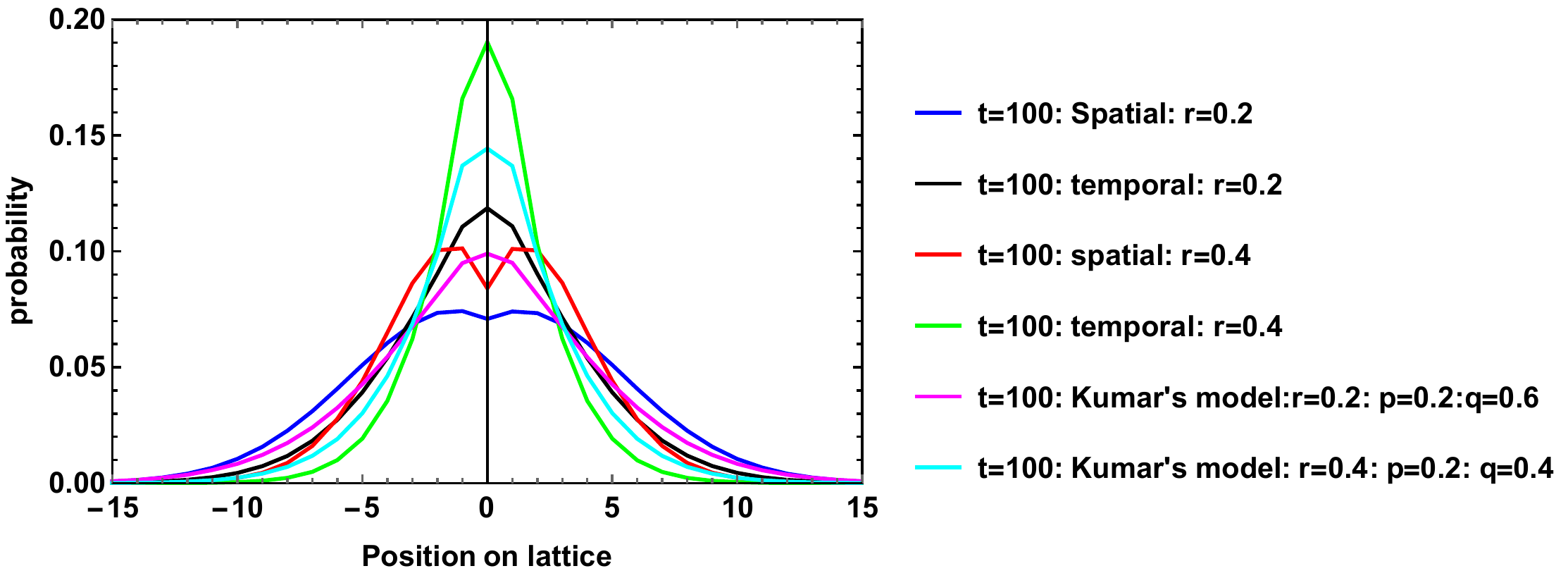}
 \caption{Figure shows probability distribution function for walk of length t=100 steps for model 2 using spatial and 
temporal correlations at stay probability r=0.2 and r=0.4}
  \label{fig8}
\end{figure}
\subsection{\label{sec:level2}Model 3}
\begin{figure}[htbp!]
\includegraphics[scale=1.0]{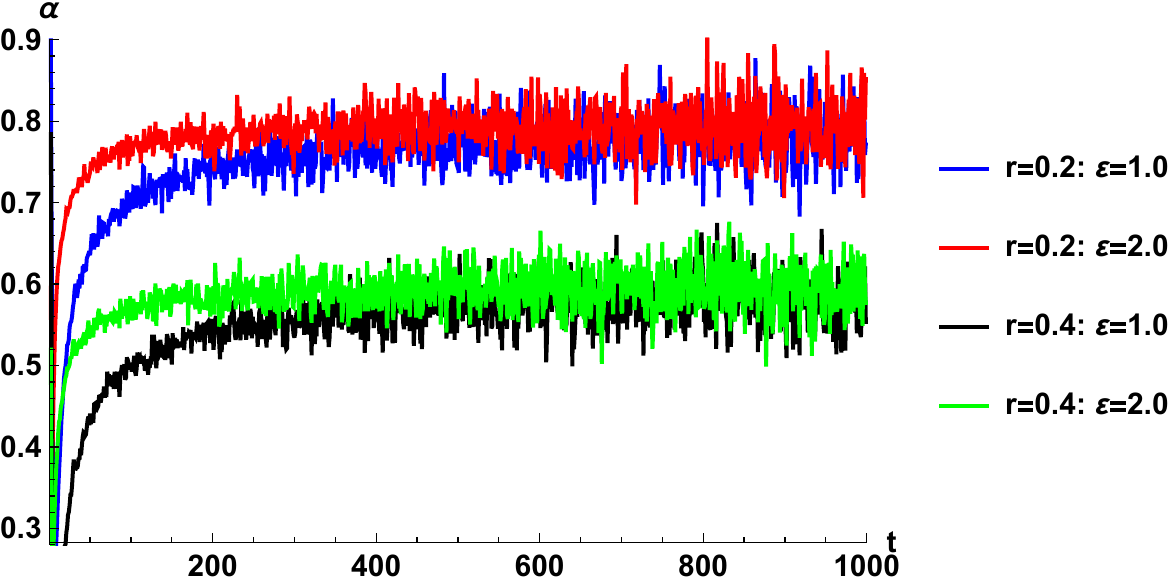}
 \caption{Figure shows variation of diffusion exponent ($\alpha$) for model 3 plotted 
against time (t) for stay probability r=0.2 and r=0.4 with heterogeneity 
parameter $\epsilon=1.0$ and $\epsilon=2.0$.}
  \label{fig9}
\end{figure}

\begin{figure}[htbp!]
\includegraphics[scale=1.0]{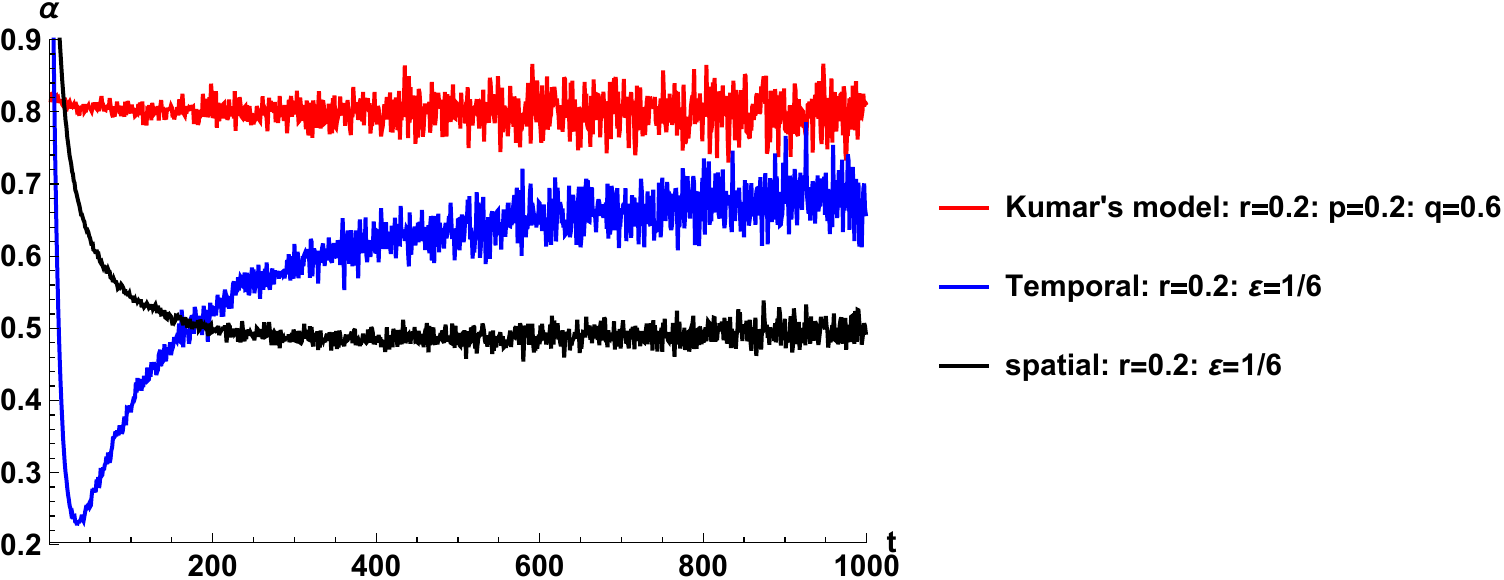}
 \caption{Figure shows variation of diffusion exponent ($\alpha$) for model 3 plotted 
against time (t) at $\epsilon=1/6$ for both spatial and temporal correlated walk 
at stay probability r=0.2.}
  \label{fig11}
\end{figure}

The MC simulations have been performed using p given in Eq. \ref{EQmod3}, where 
$\epsilon$ is a parameter that determines change in the value of p with time. 
MSD and $\alpha$ are calculated from the simulation. Figure \ref{fig9} shows 
changes in $\alpha$ against time t as obtained from MC simulations at 
$\epsilon=1$ and $\epsilon=2$. For a fixed value of r, dynamics changes 
significantly with $\epsilon$. However, this change is significant only over the 
transient dynamics. The diffusion exponent ($\alpha$) approaches the value 1-r 
asymptotically for the given values of $\epsilon$ , as shown in figure. 
For small values of $\epsilon$ , it takes longer time to reach the constant 
$\alpha$ value. For all values of $\epsilon$ considered, model 3, similar to 
model 2, gives only subdiffusion with exponent, $\alpha=1-r$. This is due to the 
decrease in the value of `p' with increase in time and hence results in increase 
in the value of `q'. The subdiffuive dynamics with exponent exponent $1-r$ is in 
accordance with the Kumar's model when $p-q<0.5$.\\
We next consider spatial correlation for model 3. The corresponding probability 
p for spatial correlated walk is given as\\
\begin{eqnarray}
\label{old34}
p=(1+(x(t-1)-x(k))^2)^{(-\epsilon)}(1-r)
\end{eqnarray}
Using Eq. \ref{old34} for the biasing probability `$p$' we performed MC 
simulations to calculate MSD and diffusion exponent ($\alpha$ ) for walks of 
lengths up to 1000 steps.\\
Figure \ref{fig11} shows comparison of $\alpha$ values for spatial and temporal 
correlations at $\epsilon=1/6, r=0.2$. For temporal correlated walk, the 
diffusion exponent ($\alpha$) first decreases and then increases till it attains 
diffusion exponent $\sim 0.7$ within the given time interval of time(however at long 
time as mentioned it goes to 1-r). However, for spatial correlated walk, 
diffusion exponent decrease monotonically till it reaches a constant value $\sim 0.5$ . The exponent value in case of spatially correlated walk is found to be 
lower than the temporal correlated walk. However, for spatial correlated walk 
the value of p is larger than the temporally correlated walk. For spatial 
correlation, it is expected that the change in value of p is slower in case of 
temporally correlated walk which is due the fact that number of distinct 
positions covered (from time $1$ to $t-1$) is always less than the number of time 
steps covered (history positions are very few in comparison to time of history 
which is from $1$ to $t-1$). The slow rate of decrease of p is also responsible for 
slow rate of increase of MSD with time. The slow rate of change of MSD suggests 
higher subdiffusive behavior in spatial correlated walk in comparison to 
temporal correlated walk. Figure \ref{fig11} also shows comparison of diffusion 
exponent ($\alpha$) from the given model and Kumar's model at stay probability 
of 0.2. From the figure, it can be seen for same value of $\epsilon$ spatial correlation 
gives lower value of $\alpha$ in comparison to temporal correlation and Kumar's model 
at stay probability of 0.2.
\subsection*{\label{sec:level2}Summary of the models}
The models discussed in this work for the probability of following the past step 
hold importance in the dynamics of a particle. The model 1 describes the 
dynamics when the environment induced temporal correlations are such that the steps which are farther are followed with larger probability than the steps closer to the 
current time. This leads to the motion with all three types of dynamics, normal 
diffusion, superdiffusion, and subdiffusion and shows a rich phase diagram. 
However, model 2, where the distant steps have a lower probability to be 
followed than the ones closer to the current step, always shows subdiffusive 
behavior and the average position of the particle remains unchanged at long 
times. The effect of environment has also been introduced in model 2 by 
including the correlation between the current position and the position of 
particle at randomly selected past step. This can implicitly account for 
disordered environment providing correlation between positions of particle. The 
qualitative behavior of model 3 is similar model 2. In the model 3 also, the 
distant history is followed with less probability than the immediate one which 
may be the reason that for both models 2 and 3 we always get subdiffusion. Like 
model 2, the effect of spatial correlation has also been determined for model 3.
For temporal correlation in model 3, the local heterogeneity is not of 
consequence for large time dynamics and only the average crowding (r) dictates 
the diffusive dynamics.
\section{\label{sec:level1}CONCLUSION}
In the current work, we proposed random walk based models to understand 
diffusion in crowded and heterogeneous environment. The crowding and 
heterogeneity of the environments have been implicitly considered by introducing 
biasing and temporal and spatial correlations in between the past and present 
moves. The probability of particle following the past steps and their dependence 
on time and space relates the motion of particle to the environment in which 
particle undergoes diffusive motion. The models discussed in our study can 
produce both normal and anomalous (subdiffusive and superdiffusive) diffusion 
with different set of parameters incorporated to account for memory that is 
induced due to heterogeneity of the environment. The Gaussian correlation 
induced due to environment does not lead to superdiffusion while a power law 
correlation may or may not give rise to superdiffusion depending on the type of 
the power law behavior as in model 1 and 3 introduced in the study. The models 
developed in our study can be utilized to reproduce subdiffusion observed in the 
various biological processes that involve motion of a particle in a crowded 
complex environment. The complexity of the environment can be incorporated in 
the time and/or spatial dependence of the probability of following a selected 
past step.\\
Using three models, we can implicitly relate to the heterogeneity of the 
environment depending on how well particle remembers and follows the history. 
The correlation and the memory of history related problems have its significance 
in the problems related to stochastic modeling of animal movement. Large number 
of studies are there in the animal movement to specific regions is based on the 
history of how strongly they remember their past movements which depends on 
factors like food, environment, safely etc. 
\cite{worton1987review,smouse2010stochastic,horne2007analyzing}. Depending on 
the how strongly particle remembers near and far history the three models can be 
used in different cases. For the system in which the particle has strong memory 
of far history, model 1 can be employed. However, for the systems for which 
particle remembers near history more strongly, then model 2 or 3 can be used.
\nocite{*}
\begin{acknowledgments}
This work has been funded by a DST grant awarded to P.B. (SB/S1/PC-048/2013) and 
also by a UPE grant awarded to P.B. from Jawaharlal Nehru University. S.H. was 
funded by the Maulana Azad fellowship from University Grant Commission (UGC). 
U.H acknowledges financial support from Indian Institute of Science, Bangalore, 
India. Mr. Ram Nayan Verma is acknowledged for the initial work related to this 
paper.
\end{acknowledgments}
\bibliography{file1}

\end{document}